\documentclass[reprint]{revtex4-1}
\usepackage{preamble}

\begin{document}

\title{A relativistic gravity train}
\date{\today}
\author{Edward Parker}
\email{tparker@physics.ucsb.edu}
\affiliation{Department of Physics, University of California, Santa Barbara, CA 93106}

\begin{abstract}
A nonrelativistic particle released from rest at the edge of a ball of uniform charge density or mass density oscillates with simple harmonic motion.  We consider the relativistic generalizations of these situations where the particle can attain speeds arbitrarily close to the speed of light; generalizing the electrostatic and gravitational cases requires special and general relativity, respectively.  We find exact closed-form relations between the position, proper time, and coordinate time in both cases, and find that they are no longer harmonic, with oscillation periods that depend on the amplitude.  In the highly relativistic limit of both cases, the particle spends almost all of its proper time near the turning points, but almost all of the coordinate time moving through the bulk of the ball.  Buchdahl's theorem imposes nontrivial constraints on the general-relativistic case, as a ball of given density can only attain a finite maximum radius before collapsing into a black hole.  This article is intended to be pedagogical, and should be accessible to those who have taken an undergraduate course in general relativity.
\end{abstract}

\maketitle

\section{Introduction}

A classic problem in elementary gravitation is the ``gravity train'' \cite{KK}: calculating the trajectory of a test particle under the gravitational influence of a ball with uniform mass density (often taken to be the Earth) as the particle falls through the ball's diameter.  In the nonrelativistic limit, the force that the test particle feels while inside the ball is exactly that of a simple harmonic oscillator, so the particle oscillates sinusoidally with a period independent of its initial radius or velocity, as long as it stays inside the ball over its entire trajectory.  As a result, the period of oscillation (about 84 minutes, in the case of the Earth) turns out to depend only on the ball's density, and not on any extensive quantities such as its radius or total mass.  (In fact, the motion of a test particle confined by frictionless forces to move through \emph{any} ``tunnel through the Earth'' -- that is, any chord of the ball -- is also precisely harmonic, with a frequency that does not depend on the length of the coord \cite{Cooper}!  But in this article, we will only consider trajectories along a diameter.)

A natural extension to this well-studied problem is to incorporate relativistic effects.  In Section~\ref{Nonrel}, we briefly summarize the nonrelativistic situation, which can be equivalently treated in terms of balls of uniform charge density or mass density.  We then consider two natural ways to generalize the nonrelativistic problem into the relativistic setting.  In Section~\ref{Special}, we generalize the electrostatic case of the ball of uniform charge to allow the test particle to move arbitrarily close to the speed of light $c$, which requires incorporating the effects of special relativity (SR).  We find the remarkable result that, up to constants, the exact relativistic trajectory is simply the sine of the trajectory of a simple pendulum outside of the small-angle regime.  We also relate this situation to the famous ``twin paradox'' of special relativity.  In Section~\ref{General}, we generalize the gravitational case of the ball with uniform mass to again allow the test particle to move arbitrarily close to the speed of light, which requires incorporating the effects of general relativity (GR).  Remarkably, the GR situation is ``less relativistic'' than the SR situation, in the sense that the ball cannot become dense enough to cause huge relativistic effects without collapsing into a black hole.  We also find that Newton's shell theorem still holds in the SR case, but is violated in the GR case.  In both cases, the nonrelativistic simple harmonic oscillator becomes a nonlinear, and much more complicated, relativistic harmonic oscillator.  The particle's trajectory therefore acquires a complicated dependence on its amplitude, but remarkably, the position, proper time, and coordinate time can all be exactly related to each other in terms of elliptic integrals.  In both cases, a highly relativistic particle spends almost all of its \emph{proper} time near the turning points, but almost all of its \emph{coordinate} time moving through the ball's bulk.  In Section \ref{Conclusion}, we conclude by summarizing the key similarities and differences between the SR and GR cases.

\section{Warm-up: nonrelativistic case \label{Nonrel}}

The nonrelativistic case is a standard problem in Newtonian gravitation; we will assume that the reader is familiar with it and only briefly summarize the key results.  Consider a ball of radius $R$ and uniform mass density $\rho_m$.  Since the mass density is spherically symmetric, Newton's shell theorem gives that a test particle of mass $m$ at radius $r$ only experiences a net gravitational pull from the mass $M(r) = (4/3) \pi r^3 \rho_m$ at a smaller radius than the test particle, and experiences the same force as if that mass were all concentrated at the origin.  So at point $\bm{r}$, it experiences a net force
\[
\bm{F}(\bm{r}) = -\frac{G M(r)\, m}{r^2} \hat{\bm{r}},
\]
where $G$ is Newton's constant and the unit vector $\hat{\bm{r}} := \bm{r}/r$, leading to the one-dimension equation of motion
\beq \label{NREOM}
\frac{d^2x}{dt^2} = -\omega^2 x,
\eeq
where $x$ is the test particle's position along the diameter, $t$ is time, and the frequency of oscillation
\beq \label{omegam}
\omega := \sqrt{\frac{4}{3} \pi G \rho_m}.
\eeq
It is easy to show that this is the same period as the orbit of a free-falling satellite orbiting the sphere at exactly its radius $R$; in fact, the trajectory of the test particle through the sphere's diameter is simply the projection of such a satellite's orbit onto any diameter of that orbit.

We find that the (nonrelativistic) period of oscillation
\[
T_\text{NR} = \frac{2 \pi}{\omega}
\]
depends only on the sphere's density, not on its size.  The Earth has an average density of about $\rho_m = 5515 \text{ kg/m}^3$, so if we neglect all forces except gravity, the period to fall through the center of the Earth and return is about $84$ minutes.  (The Sun is about $1/4$ as dense as the Earth, so in the same approximation, it takes about twice as long to fall through the Sun and return.)

Most students are surprised when they first learn that the oscillation period does not depend on the sphere's radius, but this result follows simply from dimensional analysis.  The only relevant independent quantities are $G$, $\rho_m$, $R$, and $m$.  The equivalence principle requires all particles to follow the same free-falling trajectories regardless of their mass, so the period cannot depend on $m$.  In terms of the fundamental units of mass $M$, length $L$, and time $T$, the remaining quantities have dimensions $[G] = L^3 / (M T^2)$, $[\rho_m] = M / L^3$, and $[R] = L$.  Since only $G$ contains a time scale, we find that $\omega_m \propto \sqrt{G \rho_m}$ cannot depend on $R$ or any other extensive properties of the sphere, making the Hooke's-law gravitational force very natural.

The case of a test charge $q$ of mass $m$ electrically attracted to an insulating sphere of radius $R$ and uniform electric charge density $\rho_c$ is essentially identical.  At point $\bm{r}$, the test charge feels an electrostatic force
\[
\bm{F}(\bm{r}) = \frac{Q(r)\, q}{r^2} \hat{\bm{r}},
\]
where $Q(r) = (4/3) \pi r^3 \rho_c$ is the total charge lying closer to the origin than the test charge.  (We work in CGS units to simplify the relativistic generalization below.)  If $q$ and $\rho_c$ have opposite sign, then the one-dimensional equation of motion is again given by \eqref{NREOM}, but with
\beq \label{omegac}
\omega := \sqrt{\frac{4 \pi |\rho_c\, q|}{3 m}}.
\eeq

Although the cases of the spheres of uniform mass and uniform charge are mathematically identical in the nonrelativistic limit, they have different relativistic generalizations, as there is an equivalence principle for gravity but not for electromagnetism.  One way to think about the difference is that electric charge density $\rho_c$ transforms under Lorentz transformations as the $0$-component of a four-vector $J^\mu$, while mass density transforms as the $00$-component of a rank-$2$ tensor $T_{\mu \nu}$.  Roughly speaking, this is because under a Lorentz boost, charge density is multiplied by one factor of $\gamma := 1 / \sqrt{1 - (v^2/c^2)}$ due to the Lorentz contraction of space in the boosted direction, while mass density is multiplied by two factors of $\gamma$: one due to the Lorentz contraction of space and another due to relativistic mass dilation.

\section{Special relativity: sphere of uniform charge \label{Special}}

The relativistic Lorentz force law for a point charge $q$ of mass $m$ is \cite{Jackson}
\[
F^\mu = m \frac{d^2 x^\mu}{d\tau^2} = \frac{q}{c} F^{\mu \nu} U_\nu \\
\]
where $F^\mu$ is the four-force, $x^\mu$ is the four-vector $(ct, {\bf r})$, $\tau$ is the proper time, $F^{\mu \nu}$ is the electromagnetic tensor, and $U^\mu$ is the four-velocity $dx^\mu/d\tau$ (we use the metric sign convention $(-,+,+,+)$).  In our case, there is an electric field $\bm{E}$ but no magnetic field, and we can assume without loss of generality that the particle moves along the $x$-axis, so this equation reduces to
\begin{align}
m \frac{d^2 x^0}{d\tau^2} &= \frac{q E^x}{c} U^1 \label{EOM0} \\
m \frac{d^2 x^1}{d\tau^2} &= \frac{q E^x}{c} U^0. \label{EOM1}
\end{align}
(In the particle's frame, there will be magnetic fields induced by the apparent motion of the ball of charge, but not along the particle's axis of motion.)  Equation \eqref{EOM0} can be equivalently written in many different ways as
\[
\frac{dp^0}{d \tau} = F^0 = \frac{\gamma}{c} \bm{F} \cdot \bm{v} = \frac{\bm{F} \cdot \bm{w}}{c} = \frac{q}{c} \gamma \bm{v} \cdot \bm{E} = \frac{q}{c} \bm{w} \cdot \bm{E},
\]
where $p^0$ is the charged particle's kinetic plus rest energy $\gamma m c^2$, $\bm{F}$ is the (three-vector) force, $\bm{v}$ is the coordinate velocity $d\bm{x}/dt$, and $\bm{w}$ is the proper velocity $d\bm{x}/d\tau = \gamma \bm{v}$.  It is simply a statement of the relativistic work-energy theorem that the change in the particle's mechanical energy equals the net work done on it by applied forces.  Equation \eqref{EOM1} is the particle's equation of motion, parameterized by its proper time.

The integral form of Gauss's law gives that a sphere of radius $R$ and uniform charge density $\rho_c$ produces an electrostatic field
\[
\bm{E}(\bm{r}) = \frac{4}{3} \pi \rho_c\, r\, \hat{\bm{r}}
\]
for $r \leq R$, so in this region the equation of motion \eqref{EOM1} becomes
\beq \label{SREOM}
m \frac{d^2 x^1}{d\tau^2} = -m \omega^2\, x^1 \frac{U^0}{c},
\eeq
where $\omega$ is defined in \eqref{omegac}.  (In this article, the letter $\omega$ always denotes a \emph{non}relativistic, purely sinusoidal frequency.)  We see that the system corresponds to a special-relativistic generalization of the harmonic oscillator.  This problem has been studied sporadically, including quite recently; \cite{Moreau} gives a good pedagogical treatment, \cite{Harvey, Babusci} give more advanced applications, and \cite{delaFuente} discusses a different special-relativistic generalization of the nonrelativistic oscillator.  However, most of the articles that the author could find use either advanced mathematical techniques or the Lagrangian formulation of classical mechanics, which is notoriously cumbersome and difficult to make self-consistent in the case of relativistic point particles (unlike relativistic fields) \cite{Goldstein}.  Instead, we give an alternate derivation of the relativistic equations of motion that follows directly from the relativistic Lorentz force law.  In the author's opinion, this derivation is simpler and more conceptually straightforward than any of those in the previous literature.

Multiplying \eqref{EOM0} by $c$ gives than inside the sphere,
\[
mc\, \frac{d U^0}{d\tau} = -m \omega^2\, x^1 \frac{dx^1}{d\tau},
\]
which can be integrated to
\[
mc\, U^0 + \frac{1}{2} m \omega^2 \left( x^1 \right)^2 = E = mc^2 + \frac{1}{2} m (\omega R)^2,
\]
where the total (rest plus kinetic plus potential) energy $E$ is constant and given by the radius $R$ of the point of release.  Solving for $U^0$ and plugging into \eqref{SREOM} gives
\[
m \frac{d^2 x^1}{d\tau^2} = -\frac{\omega^2 E}{c^2} x^1 + \frac{1}{2} m \frac{\omega^4}{c^2} \left( x^1 \right)^3.
\]
$1/\omega$ and $c/\omega$ provide the natural time and length scales, respectively, so we define
\beq \label{nondim}
\tilde{\tau} := \omega \tau, \qquad \tilde{x} := \frac{\omega x^1}{c}
\eeq
and nondimensionalize this equation to
\beq \label{ODE}
\frac{d^2 \tilde{x}}{d \tilde{\tau}^2} = - \tilde{E} \tilde{x} + \frac{1}{2} \tilde{x}^3.
\eeq
The dimensionless ratio
\beq \label{SRE}
\tilde{E} := \frac{E}{m c^2} = \gamma + \frac{U}{m c^2}
\eeq
has the constant value
\begin{gather}
\tilde{E} = 1 + \frac{1}{2} \tilde{R}^2, \label{Etilde} \\
\tilde{R} := \frac{\omega R}{c} = \frac{v_\text{NRmax}}{c} = \sqrt{\frac{q\, Q / R}{m c^2}}, \label{Rtilde}
\end{gather}
where $\tilde{R}$ is the nondimensionalized value of $R$, $v_\text{NRmax}$ is the particle's maximum speed in the nonrelativistic limit, and $Q$ is the ball's total charge.  The parameters $\tilde{E}$ and
\[
\tilde{R} = \sqrt{2 \left( \tilde{E} - 1 \right)}
\]
both measure how relativistic the system is.  We will also define the quantity
\beq \label{Rplus}
\tilde{R}_+ := \sqrt{2 \left( \tilde{E} + 1 \right)} = \sqrt{4 + \tilde{R}^2}
\eeq
for later convenience, as it will come up often.

\eqref{ODE} is a special case of a more general ordinary differential equation known as the Duffing equation.  We see that relativistic effects contribute an anharmonic term to the nonrelativistic harmonic oscillator, as we might have expected -- but perhaps surprisingly, even arbitrarily relativistic corrections can be \emph{exactly} captured by the simplest possible anharmonic modification of the harmonic oscillator (when the equation of motion is parameterized by proper time).  Remarkably, the equation of motion \eqref{ODE} is simpler than, for example, the equation of motion
\beq \label{pendEOM}
\frac{d^2\theta}{d\tilde{t}^2} = -\sin \theta
\eeq
for a simple pendulum (nondimensionalized by the small-angle frequency), in that the dependence on $\tilde{x}$ is only polynomial rather than trigonometric.

As a sanity check, let us make sure that \eqref{ODE} has the correct nonrelativistic limit by expanding to order $(v/c)^2$ in the particle's velocity.  $\tilde{x} \sim \tilde{R} \sim v / c$, so we can neglect the $\tilde{x}^3$ term in \eqref{ODE}.  Now $\tilde{E} = \gamma + U / (m c^2)$, where the potential energy $U$ has a maximum value of $(1/2) m v_\text{NRmax}^2$ in the nonrelativistic limit, so $\tilde{E} = 1 + o((v/c)^2)$.  But $\tilde{E}$ is multiplied by $\tilde{x} \sim v/c$ in \eqref{ODE}, so we can neglect the $o((v/c)^2)$ term and let $\tilde{E} \approx 1$.  Converting the derivative with respect to the proper time $\tau$ to the coordinate time $t$ only adds higher-order corrections, and restoring the proper units indeed results in the nonrelativistic equation of motion \eqref{NREOM}.

\begin{figure}
\includegraphics[width=\columnwidth]{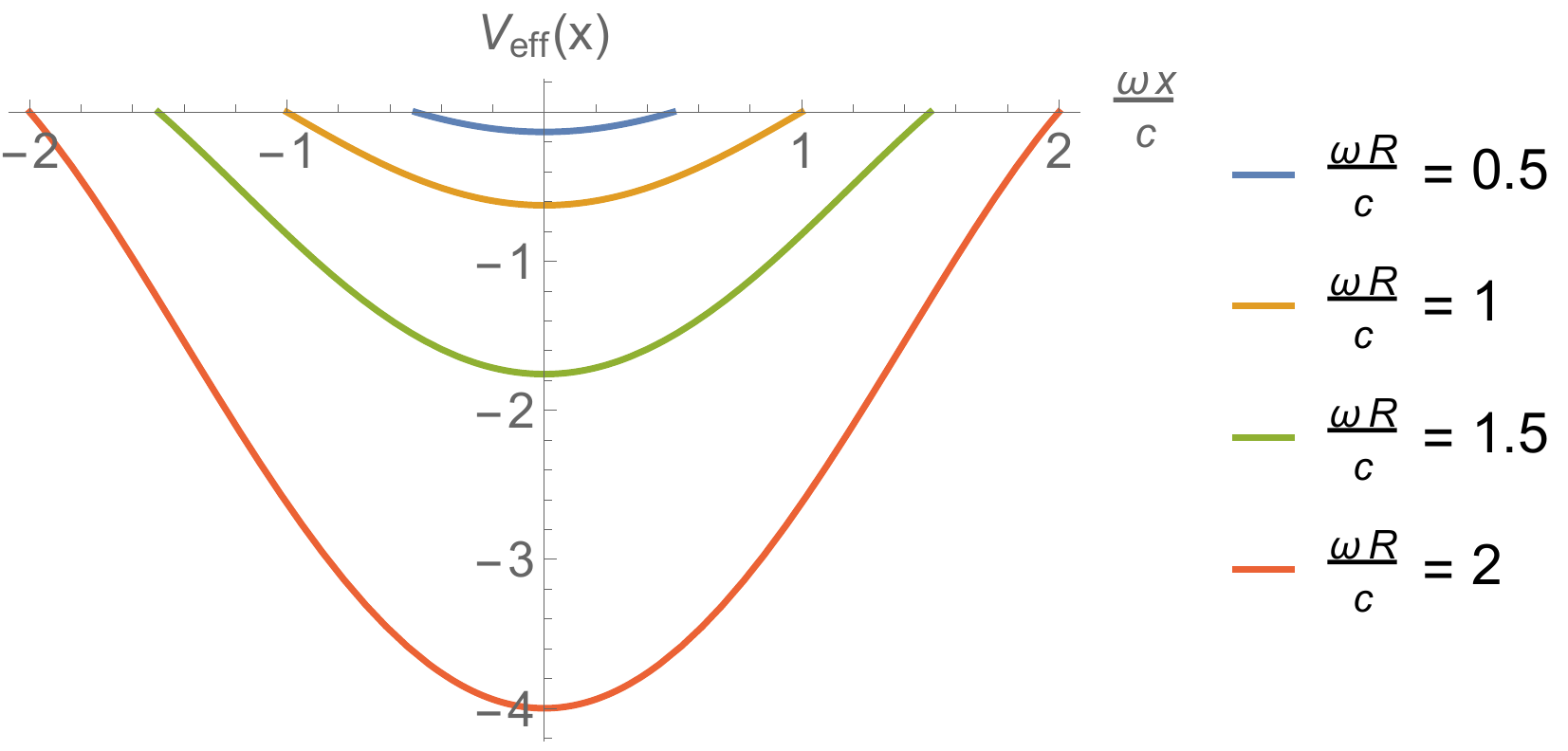}
\caption{\label{SRVeffs} Effective potential \eqref{SRVeff} for various values of $\tilde{R}$.  Each potential has been separately shifted by a constant so that the particle has zero total energy.  For $\tilde{R} < 1$ the potential is qualitatively harmonic, but for $\tilde{R} > 1$ it curves downward at the endpoints, resulting in a \emph{weaker} effective force at the ball's outer edge than slightly into the bulk.}
\end{figure}

Equation \eqref{ODE} is formally identical to the equation of motion of a nonrelativistic particle in an anharmonic effective potential per unit mass
\beq \label{SRVeff}
V_\text{eff}(\tilde{x}) = \frac{1}{2} \tilde{E} \tilde{x}^2 - \frac{1}{8} \tilde{x}^4,
\eeq
so we can solve it by taking advantage of an integral of motion that equivalent to the total energy.  Multiplying \eqref{ODE} by the integrating factor $d\tilde{x} / d\tilde{\tau}$ and integrating with respect to $\tau$ gives that the quantity
\beq \label{IOM}
\frac{1}{2} \left( \frac{d\tilde{x}}{d\tilde{\tau}} \right)^2 + V_\text{eff}(\tilde{x}) = \text{const.},
\eeq
which formally corresponds to the conservation of the equivalent nonrelativistic particle's total energy per unit mass.  $d\tilde{x}/d\tilde{\tau} = (1/c)\, dx/d\tau$, so by considering the stationary point at $\tilde{x} = \tilde{R}$, we find that the constant equals $V_\text{eff} \big( \tilde{R} \big)$ or
\[
\frac{1}{2} \tilde{R}^2 + \frac{1}{8} \tilde{R}^4 = \frac{1}{2} \left( \tilde{E} - 1 \right) \left( \tilde{E} + 1 \right) = \frac{1}{8} \tilde{R}^2 \tilde{R}_+^2.
\]
The effective potential is plotted in Fig.~\ref{SRVeffs}.  The anharmonic term weakens the effective force on the particle at large $|\tilde{x}|$ relative to a Hooke's-law force (although interestingly, if we restore the dimensionful quantities, we find that at the turning point $x = R$ the four-acceleration $d^2x / d\tau^2$ equals the equivalent nonrelativistic acceleration $-\omega^2 R$).  In fact, in the highly relativistic regime $\tilde{R} > 1$, near the turning points the anharmonic term dominates and the effective force on the particle actually \emph{weakens} with distance.

Equation \eqref{IOM} can be rearranged to give the nondimensionalized proper velocity
\begin{gather}
\tilde{w} := \frac{d\tilde{x}}{d\tilde{\tau}} = \pm \frac{1}{2} \sqrt{ \tilde{R}^2 \tilde{R}_+^2 - \left( \tilde{R}^2 + \tilde{R}_+^2 \right) \tilde{x}^2 + \tilde{x}^4} \label{w} \\
2 \int \frac{d\tilde{x}}{\sqrt{ \tilde{R}^2 \tilde{R}_+^2 - \left( \tilde{R}^2 + \tilde{R}_+^2 \right) \tilde{x}^2 + \tilde{x}^4}} = \pm \int d\tilde{\tau}. \nonumber
\end{gather}
The integral can be performed in closed form, although the final answer is not very enlightening:
\beq \label{tauOfx}
\frac{2}{\tilde{R}_+} F \left( \frac{\tilde{x}}{\tilde{R}}; \frac{\tilde{R}}{\tilde{R}_+} \right) = \pm \tilde{\tau},
\eeq
where $F(x; y)$ is the incomplete elliptic integral of the first kind.  (Note that $\tilde{R}$ and $\tilde{R}_+$ appear symmetrically in \eqref{w}, but \eqref{tauOfx} is not obviously symmetric in these parameters.  But we can use the identity
\beq \label{Fid}
\frac{1}{b} F \left( \frac{x}{a}; \frac{a}{b} \right) \equiv \frac{1}{a} F \left( \frac{x}{b}; \frac{b}{a} \right)
\eeq
to show that \eqref{tauOfx} is in fact symmetric in $\tilde{R}$ and $\tilde{R}_+$.)  We have fixed the integration constant by chosing the initial condition $\tilde{x}(\tilde{\tau} = 0) = 0$, where the particle starts at the center of the sphere.  From now on, we will drop $\pm$ signs in order to simplify the expressions.

\begin{figure}
\includegraphics[width=\columnwidth]{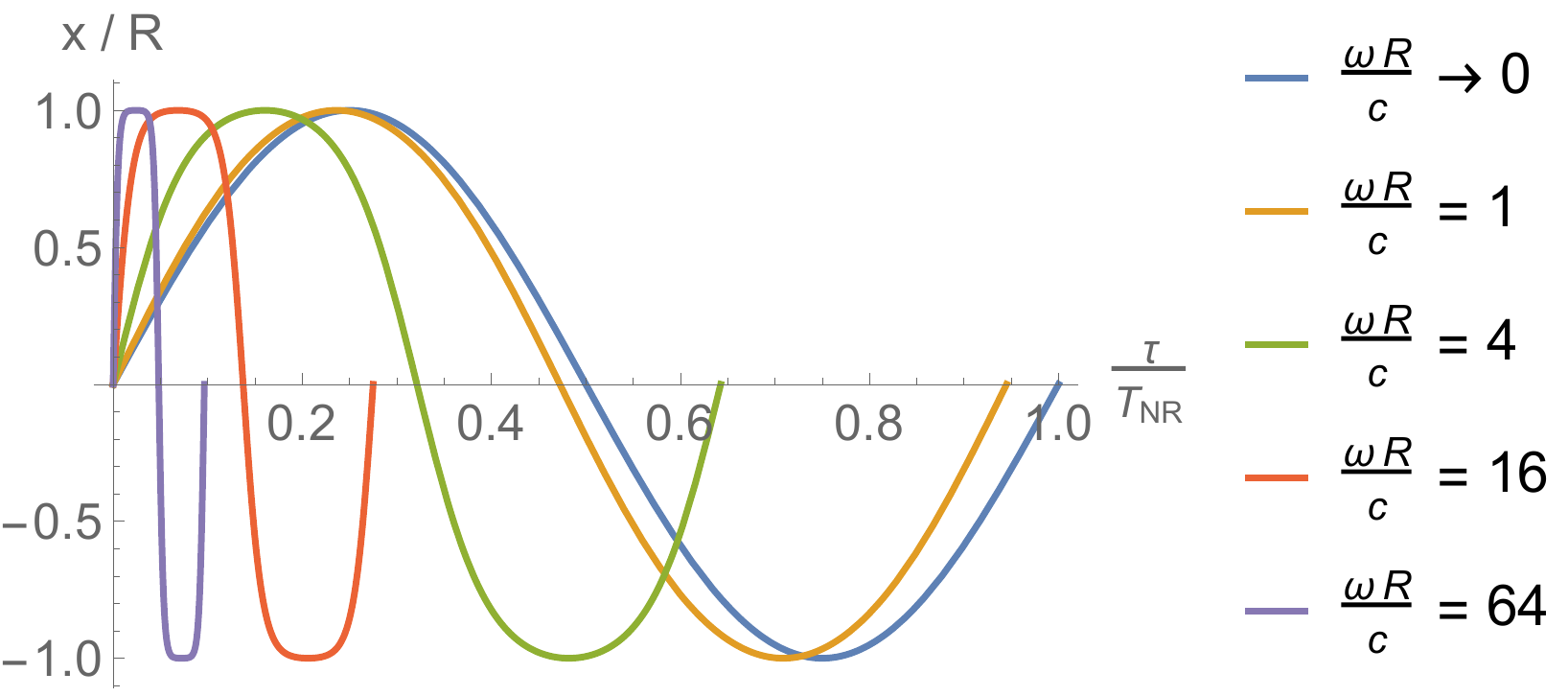}
\caption{\label{xvtau} Trajectory \eqref{xoftau} plotted against proper time over one oscillation.  Highly relativistic particles with large amplitudes $R$ observe themselves as spending most of their time near the turning points and passing through the bulk of the sphere very quickly, with an oscillation period much less than the nonrelativistic period $T_\text{NR} = 2 \pi / \omega$.  We need to go into the ultrarelativistic regime before the trajectory distortion becomes obvious, so we have normalized all the amplitudes to be the same in order to keep the less-relativistic trajectories visible.}
\end{figure}

Ref.~\cite{Moreau} claims that this expression cannot be inverted to give $\tilde{x}(\tilde{\tau})$, but in fact it can: the result is
\beq \label{xoftau}
\tilde{x} = \tilde{R}\, \mathrm{sn} \left( \frac{\tilde{R}_+}{2} \tilde{\tau}, \frac{\tilde{R}}{\tilde{R}_+} \right),
\eeq
where $\mathrm{sn}(x, y)$ is the Jacobi elliptic function.  [Again, this expression turns out to be symmetric in $\tilde{R}$ and $\tilde{R}_+$.  We caution the reader that there are many different notational conventions for the elliptic integrals and functions; in this article we use ``Jacobi's form,'' in which the first argument is the sine of the elliptic amplitude (not the amplitude itself) and the second argument is the elliptic modulus (not its square).  The elliptic moduli appearing in our expressions are sometimes formally pure imaginary, but the elliptic integrals themselves are always real.]  We can finally use \eqref{nondim}, \eqref{Rtilde}, and \eqref{Rplus} to convert this equation into the particle's trajectory $x^1(\tau)$.  The trajectories for various amplitudes $\tilde{R}$ are plotted as a function of proper time in Fig.~\ref{xvtau}.  A highly relativistic particle perceives itself as spending almost all of its time at the turning points (because the restoring force there is weakened relative to Hooke's law) and moving through the bulk of the sphere extremely quickly (because it sees the sphere as strongly Lorentz contracted, so it only needs to cross an effective diameter much shorter than $2R$).

The particle is at rest when $\tilde{x} = \tilde{R}$, and plugging this into \eqref{tauOfx} gives one-quarter of the proper-time period of oscillation
\beq
T_\text{prop} = \frac{4 \tilde{\tau} \big( \tilde{R} \big)}{\omega} = \frac{8}{\tilde{R}_+ \omega} K \left( \frac{\tilde{R}}{\tilde{R}_+} \right), \label{SRTprop}
\eeq
where the function $K(x)$ is the complete elliptic integral of the first kind (which in fact is also known as the ``quarter period'' function, precisely because it equals one-quarter of the period of the Jacobi elliptic function).  It is not at all surprising that the anharmonic $\tilde{x}^3$ term in \eqref{ODE} causes the oscillation period to depend on the amplitude; in fact, only quadratic potentials produce (nonrelativistic) oscillation periods that do not depend on the amplitude \cite{Urabe}.  The proper-time period decreases monotonically with the amplitude $\tilde{R}$ and so is always less than the nonrelativistic period.  In the particle's frame, this period shortening occurs because the ball appears Lorentz contracted, as mentioned above, while in the inertial frame, the period shortening occurs because the particle's clock appears to be running slow.  In the nonrelativistic limit $\tilde{R} \ll 1$,
\[
\frac{T_\text{prop}}{T_\text{NR}} = 1 - \frac{1}{16} \tilde{R}^2 + o\left( \tilde{R}^4 \right),
\]
and in the ultrarelativistic limit $\tilde{R} \gg 1$,
\beq \label{URTprop}
\frac{T_\text{prop}}{T_\text{NR}} = \frac{4 \ln \left( 2 \tilde{R} \right)}{\pi \tilde{R}} + o\left( \tilde{R}^{-3} \right).
\eeq
We give a simple physical argument for this form (including the prefactor) below.

As a brief aside, we note that for the simple pendulum with equation of motion \eqref{pendEOM} and initial angle $\theta_0$, the equivalent of \eqref{w} is
\beq \label{pendInt}
\int \frac{d\theta}{\sqrt{2 \left( \cos \theta - \cos \theta_0 \right)}} =\tilde{t}.
\eeq
This looks completely unlike \eqref{w}, but remarkably, the two integrals are related by an integral substitution, and \eqref{pendInt} evaluates to
\[
\csc \left( \frac{\theta_0}{2} \right) F\left( \sin \left( \frac{\theta}{2} \right); \csc \left( \frac{\theta_0}{2} \right) \right) = \tilde{t},
\]
which is quite similar to \eqref{tauOfx}.  In fact, if we let $\sin(\theta_0/2) = \tilde{R} / \tilde{R}_+$ and $t = (1/2) \tilde{R}_+ \tilde{\tau}$, then we find the remarkable result that the relativistic oscillator's trajectory $\tilde{x}(\tilde{\tau})$ (given by \eqref{xoftau}) and the exact trajectory $\theta(t)$ of the simple pendulum are related by
\[
\tilde{x}(\tilde{\tau}) = \tilde{R}_+ \sin \left( \frac{\theta(t)}{2} \right).
\]
The familiar small-angle limit of the simple pendulum -- often studied in a very first course in physics -- is therefore \emph{exactly} (not just perturbatively) mathematically dual to a nonrelativistic limit!

\begin{figure}
\includegraphics[width=\columnwidth]{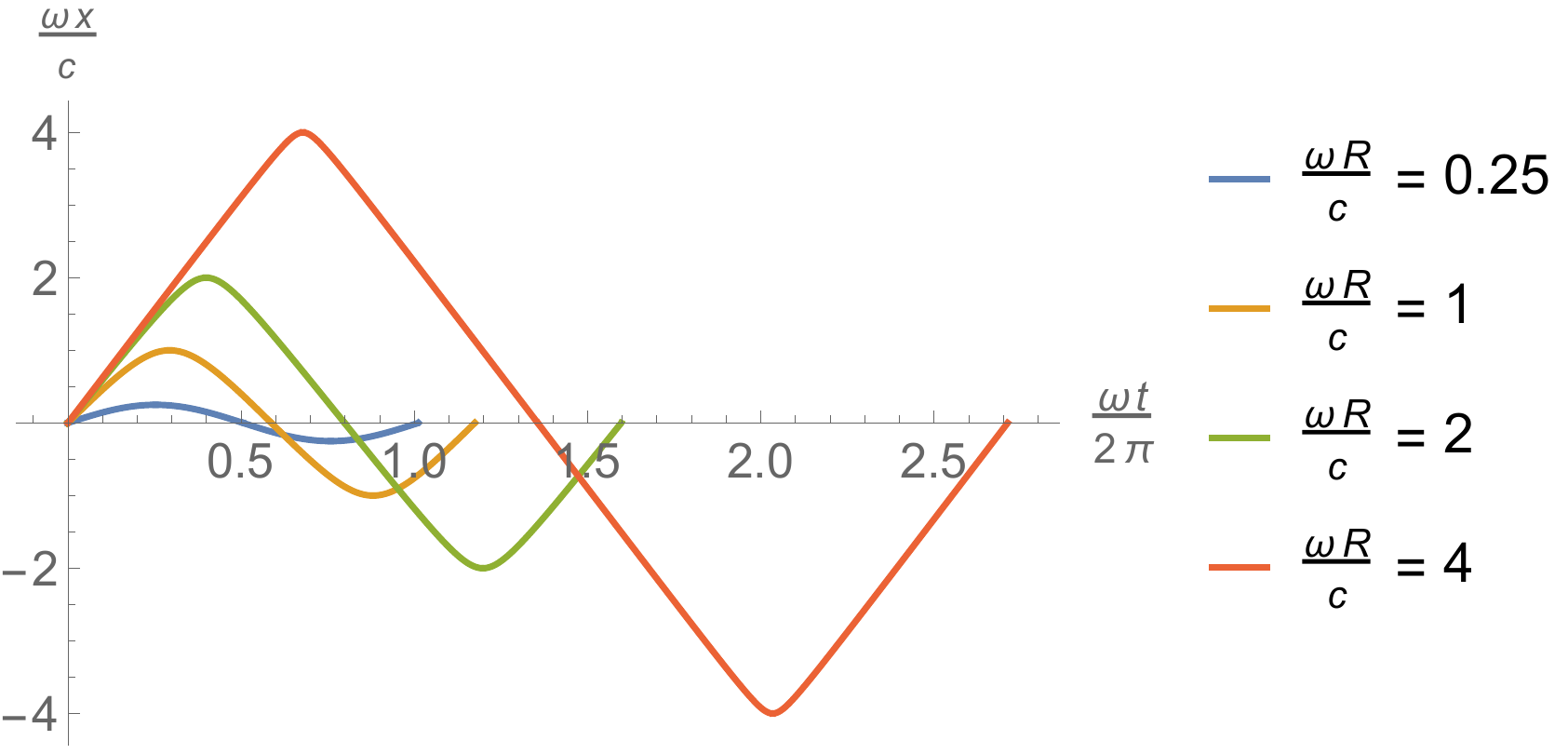}
\caption{\label{xvt} Trajectory \eqref{tofx} plotted against coordinate time over one oscillation.  Highly relativistic particles with large amplitude $R$ appear to travel just below the speed of light through the bulk of the ball and make very sharp turns at the turning points.}
\end{figure}

We can also express the particle's trajectory in terms of the coordinate time $t$ measured by an observer at rest with respect to the sphere of charge.  If we define $\tilde{t} := \omega t$, then $d\tilde{t} = \gamma\, d\tilde{\tau}$ and
\beq \label{dtdx}
\frac{d\tilde{t}}{d\tilde{x}} = \gamma \frac{d\tilde{\tau}}{d\tilde{x}} = \frac{\gamma}{\tilde{w}}.
\eeq
To express $\gamma$ in terms of the previously calculated proper velocity $\tilde{w}$, we note that the four-velocity $U^\mu = (\gamma c, \bm{w})$ and solve $U^\mu U_\mu = -c^2$ for $\gamma$ to get
\beq \label{gamma}
\gamma = \sqrt{1 + \frac{w^2}{c^2}} = \sqrt{1 + \tilde{w}^2}.
\eeq
\eqref{dtdx}, \eqref{gamma}, and \eqref{w} then give
\begin{align}
\frac{d\tilde{t}}{d\tilde{x}} &= \sqrt{1 + \frac{1}{\tilde{w}^2}} = \left( 1 - \frac{1}{\left( \tilde{E} - \frac{1}{2} x^2 \right)^2} \right)^{-\frac{1}{2}} \nonumber \\
\tilde{t} &= \tilde{R}_+\, E \left( \frac{\tilde{x}}{\tilde{R}}; \frac{\tilde{R}}{\tilde{R}_+} \right) - \tilde{\tau}, \label{tofx}
\end{align}
where $\tilde{\tau}$ is given by \eqref{tauOfx} and $E(x; y)$ is the incomplete elliptic integral of the second kind.  (Unlike $\tilde{\tau}(\tilde{x})$, $\tilde{t}(\tilde{x})$ is not symmetric in $\tilde{R}$ and $\tilde{R}_+$; the identity corresponding to \eqref{Fid} for $E(x; y)$ is
\[
\tilde{R}\, E \left( \frac{\tilde{x}}{\tilde{R}_+}; \frac{\tilde{R}_+}{\tilde{R}} \right) = \tilde{R}_+\, E \left( \frac{\tilde{x}}{\tilde{R}}; \frac{\tilde{R}}{\tilde{R}_+} \right) - \frac{4}{\tilde{R}_+} F \left( \frac{\tilde{x}}{\tilde{R}}; \frac{\tilde{R}}{\tilde{R}_+} \right),
\]
\emph{only} if $\tilde{R}_+ = \sqrt{4 + \tilde{R}}$.) We have again chosen the integration constant so that the particle begins at the origin.  $\tilde{t}(\tilde{x})$ cannot be inverted to express $\tilde{x}(\tilde{t})$ in terms of standard functions.  The trajectory is plotted as a function of coordinate time in Fig.~\ref{xvt}.  We saw above that the particle observes itself spending most of its time near the turning points, but when viewed from an inertial frame, the particle turns around extremely quickly and spends most of the time passing through the bulk of the sphere.

By considering the case $\tilde{x} = \tilde{R}$ as before, we find that the coordinate-time period of oscillation is
\beq \label{SRTcoord}
T_\text{coord} = \frac{4 \tilde{R}_+}{\omega} E \left( \frac{\tilde{R}}{\tilde{R}_+} \right) - \frac{8}{\tilde{R}_+ \omega} K \left( \frac{\tilde{R}}{\tilde{R}_+} \right),
\eeq
where $E(x)$ is the complete elliptic integral of the second kind.  The coordinate-time period increases monotonically with amplitude and so is always greater than the nonrelativistic period.  This increase occurs because the coordinate frame is inertial, so the particle's apparent speed is limited to $c$ and it cannot reach the arbitrarily high maximum speed required in order to keep the coordinate-time period constant.  In the nonrelativistic limit $\tilde{R} \ll 1$,
\[
\frac{T_\text{coord}}{T_\text{NR}} = 1 + \frac{3}{16} \tilde{R}^2 + o\left( \tilde{R}^4 \right),
\]
and in the ultrarelativistic limit $\tilde{R} \gg 1$,
\beq \label{URTcoord}
\frac{T_\text{coord}}{T_\text{NR}} = \frac{2}{\pi} \tilde{R} + o\left( \tilde{R}^{-1} \right).
\eeq
In the limit of large amplitude, the particle is so energetic that it appears to be traveling at the speed of light for almost its entire trip (as show in Fig~\ref{xvt}), so the oscillation period approaches $4 R / c = (4 / \omega) \tilde{R} = (2 / \pi) \tilde{R}\, T_\text{NR}$, in agreement with \eqref{URTcoord}.  Recall from \eqref{SRE} that the maximum Lorentz factor $\gamma$ equals $\tilde{E}$.  In the ultrarelativistic regime $\tilde{R} \gg 1$, this maximum Lorentz factor goes as $(1/2) \tilde{R}^2$ by \eqref{Etilde}; if we assume that the behavior near the origin (where relativistic effects are strongest) is the most important, then we expect that time dilation will reduce $T_\text{prop}$ to about
\[
\frac{T_\text{coord}}{\frac{1}{2} \tilde{R}^2} \sim \frac{4 }{\pi \tilde{R}}
\]
 at large $\tilde{R}$.  This indeed agrees with \eqref{URTprop}, up to a subleading logarithmic correction due to time dilation away from the ball's center.

\begin{figure}
\includegraphics[width=\columnwidth]{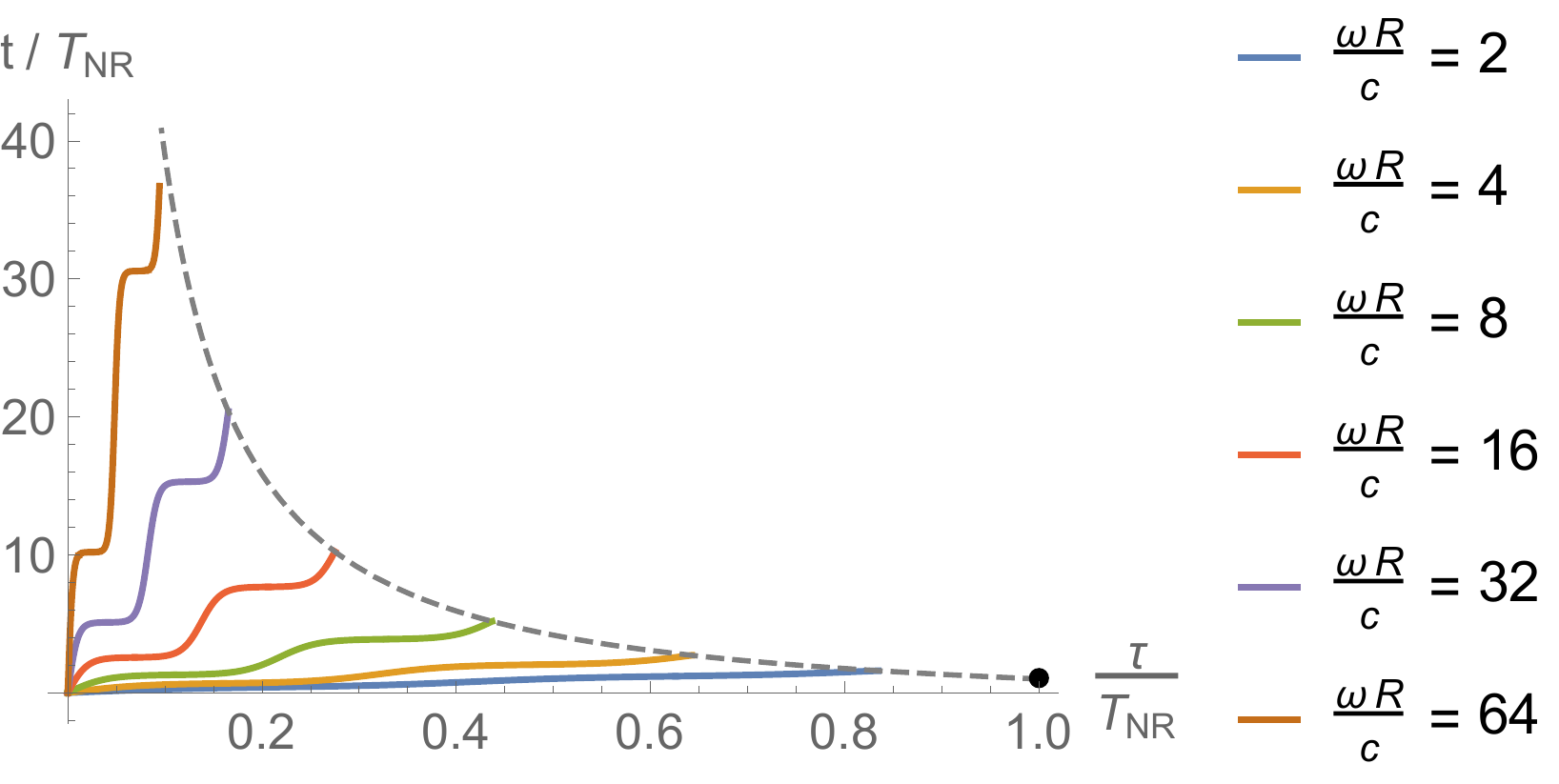}
\caption{\label{tvtau} Coordinate time plotted against proper time over one oscillation.  The dashed-line envelope comes from combining the proper- and coordinate-time periods \eqref{SRTprop} and \eqref{SRTcoord} parametrically.  The black dot indicates the nonrelativistic period.}
\end{figure}

Fig.~\ref{tvtau} shows a plot of the coordinate time against the proper time and gives another way to understand the particle's motion.  In the nonrelativistic limit, proper time and coordinate time are of course equivalent, so the plot approaches a straight line.  But in the highly relativistic limit, the curve approaches a ``staircase'' shape.  As the particle passes through the bulk of the sphere, a great deal of coordinate time elapses but very little proper time, corresponding to the nearly-vertical parts of the plot.  Near the turning points, very little coordinate time passes but a great deal of proper time does, corresponding to the nearly-horizonal parts of the plot.  Roughly speaking, the particle thinks that it is almost always turning, while the observer thinks that the particle is almost always moving straight.

In fact, the discrepancy between the proper-time and coordinate-time periods is a perfect example of the famous ``twin paradox,'' in which a person who moves relativistically away from a point and then returns ages less than a person who remains inertial at the path's endpoint.  The crucial fact that breaks the symmetry between the two observers is that one of them has to accelerate along the way in order to change direction.  In the usual presentation of the paradox, this acceleration is often thought of as brief and concentrated at the ``far end'' of the trajectory (like the highly relativistic trajectories in Fig.~\ref{xvt}), but there is no need for this to be the case, and beginning students of special relativity might find it helpful to consider the less-relativistic situations as well.

\section{General relativity: sphere of uniform mass \label{General}}

We now consider the even more complicated problem in which the ``force'' on the particle is gravitational, requiring the use of general relativity.  We first note that our argument from Sec.~\ref{Nonrel} that the period cannot depend on the amplitude by dimensional analysis breaks down in the relativistic context, because the speed of light provides a natural velocity scale.  Just as in Sec.~\ref{Special}, we can parameterize the amplitude $R$ by the dimensionless ratio
\[
\tilde{R} = \frac{\omega R}{c} = \frac{1}{c} \sqrt{\frac{G M}{R}},
\]
where $\omega$ is now defined by \eqref{omegam} and $M := (4/3) \pi R^3 \rho_m$ is the ball's mass, \emph{not} including its gravitational binding energy.  The period of oscillation can take the form $T = (2 \pi / \omega)\, f(\tilde{R})$, where $f$ is any function satisfying $f(0) = 1$.  Indeed, by the uniqueness theorem mentioned above, we can essentially guarantee that the period \emph{must} depend on the amplitude.

Before we dive into the problem, let us briefly consider in which situations GR corrections are important.  We expect them to be negligible when the particle's maximum kinetic energy $(\gamma - 1) m c^2$ is much less than its rest energy $m c^2$, and \eqref{SRE} and \eqref{Etilde} gives that (at least in the SR context) the maximum $\gamma - 1$ equals $(1/2) \tilde{R}^2$, so GR effects are negligible when $\tilde{R}^2 \ll 1$.  $\tilde{R}^2$ is approximately $7 \times 10^{-10}$ for the Earth, $2 \times 10^{-6}$ for the Sun, and $0.23$ for PSR J0348+0432, the most massive known neutron star, so any engineers currently drawing up plans for gravity trains through planets or ordinary stars can safely ignore the results of this section.  In fact, Buchdahl's theorem \cite{Carroll} requires that $\tilde{R} < 2/3$ in order for the ball to avoid collapsing into a black hole, so unlike in the SR case, the effects of GR should not change the particle's behavior by orders of magnitude.

We know that the spacetime metric $g_{\mu \nu}$ corresponding to the ball of uniform mass density must be spherically symmetric.  By choosing suitable coordinates, any spherically symmetric metric can be written in the form \cite{Carroll}
\[
ds^2 = -e^{2 \alpha(r)} c^2 dt^2 + e^{2 \beta(r)} dr^2 + r^2 d\Omega^2
\]
for some real functions $\alpha(r)$ and $\beta(r)$, where $r$ is the radial coordinate and $\Omega$ the angular coodinates.  We are only considering purely radial motion, so without loss of generality we can set $d\Omega = 0$ throughout this article.  In the case of uniform density $\rho_m$, \cite{Carroll}
\beq \label{alphabeta}
e^\alpha = \frac{3}{2} \sqrt{1 - 2 \tilde{R}^2} - \frac{1}{2} \sqrt{1 - 2 \tilde{r}^2}, \qquad e^{2 \beta} = \frac{1}{1 - 2 \tilde{r}^2}
\eeq
inside the ball, where the nondimensionalized coordinate $\tilde{r} := \omega r / c$ as usual.  (Buchdahl's theorem's requirement $\tilde{R} < 2/3$ guarantees that $e^\alpha$ and the quantities inside the square roots are positive.)  For radial motion, the metric therefore simplifies to
\beq \label{g}
ds^2 = -\left( \frac{3}{2} \sqrt{1 - 2 \tilde{R}^2} - \frac{1}{2} \sqrt{1 - 2 \tilde{r}^2} \right)^2 c^2 dt^2 + \frac{dr^2}{1 - 2 \tilde{r}^2}.
\eeq
For clarity, we continue labeling the coordinates as $ct$ and $r$, rather than as $x^0$ and $x^1$ as in Sec.~\ref{Special}.  Outside the ball, the metric is given by the Schwarzchild metric, but we will not be concerned with this region.

The simplest trajectory to consider is that of a massless particle, which we will refer to as a ``light ray'' to distinguish it from the more usual massive case.  (Of course, it need not have anything to do with electromagnetism; we are simply considering null geodesics.)  Such a trajectory cannot be bounded, or the ball would be a black hole by definition.  We will calculate the trajectory $\tilde{r}(\tilde{t})$ to verify this.  Note that since physical forces (as opposed to gravity) cannot act on a classical massless particle, any null trajectory in $(1 + 1)$ dimensions is automatically a geodesic.  Therefore, we do not even need the geodesic equation, and can simply consider a general radial null trajectory:
\begin{gather}
dx_\mu\, dx^\mu = e^{2 \beta} dr^2 - e^{2 \alpha}\, c^2 dt^2 = 0 \nonumber \\
\frac{d\tilde{r}}{d\tilde{t}} = \frac{1}{c} \frac{dr}{dt} = e^{\alpha - \beta} = \left( \frac{3}{2} \sqrt{\left( 1 - 2 \tilde{r}^2 \right) \big( 1 - 2 \tilde{R}^2 \big)} + \tilde{r}^2 - \frac{1}{2} \right) \label{masslessdrdt} \\
\int \frac{d\tilde{r}}{\frac{3}{2} \sqrt{\left( 1 - 2 \tilde{r}^2 \right) \big( 1 - 2 \tilde{R}^2 \big)} + \tilde{r}^2 - \frac{1}{2}} = \int d\tilde{t} \nonumber
\end{gather}
\begin{multline}
\frac{\tilde{r} \sqrt{4 - 9 \tilde{R}^2} \left(1 + 3 \sqrt{\left(1 - 2 \tilde{r}^2 \right) \left(1 - 2 \tilde{R}^2 \right)} \right)}{4 - 9 \tilde{R}^2 - 9 \tilde{r}^2 \left( 1 - 2 \tilde{R}^2 \right)} \\
= \tan \left( \sqrt{4 - 9 \tilde{R}^2}\, \tilde{t} \right), \label{masslessroft}
\end{multline}
where we have chosen to start the light ray at the origin, as in Sec.~\ref{Special}.  This can be rearranged to a quartic equation in $\tilde{r}$, so $\tilde{r}(\tilde{t})$ can in principle be expressed in terms of roots of $\tan \left( \sqrt{4 - 9 \tilde{R}^2}\, \tilde{t} \right)$, but the resulting expression is not illuminating.

\begin{figure}
\includegraphics[width=\columnwidth]{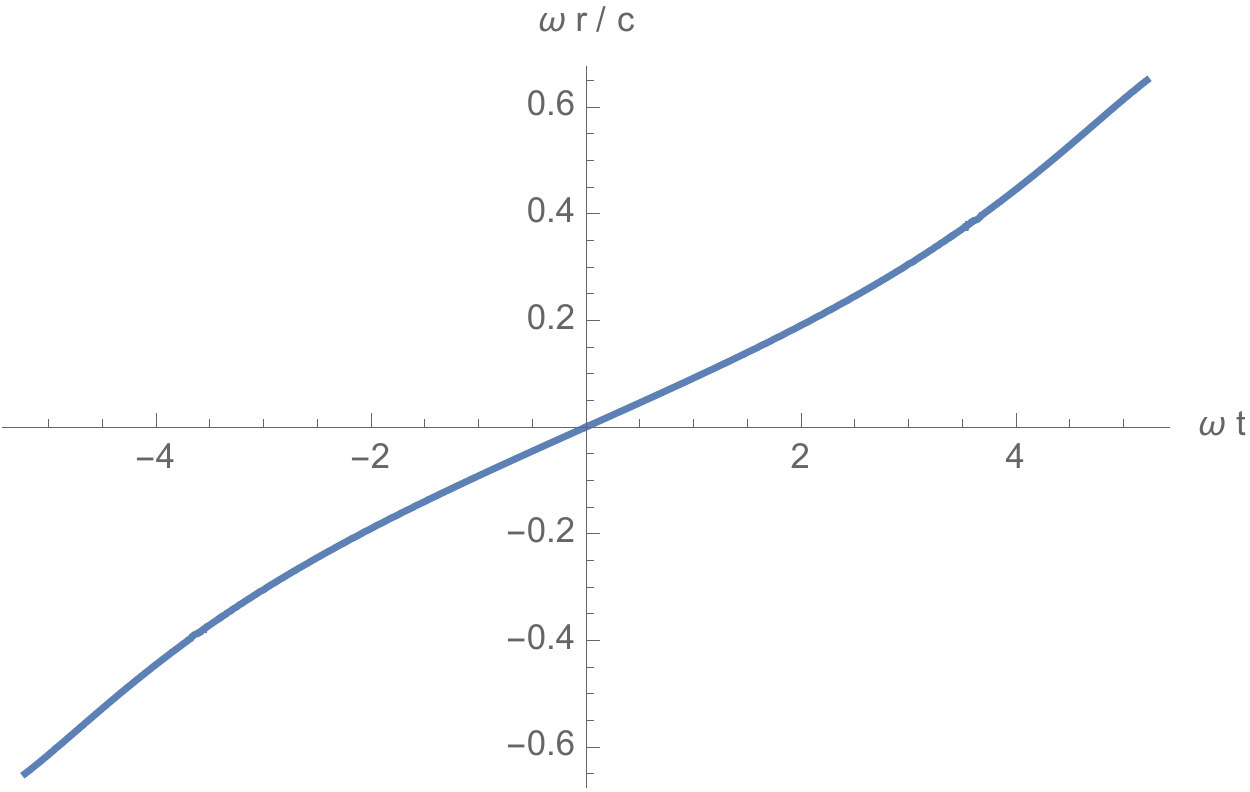}
\caption{\label{Massless} Null geodesic trajectory \eqref{masslessroft} plotted against coordinate time for a ball with $\tilde{R} = 0.65$ very close to the Buchdahl bound of $2/3$.  Note the scale of the axes: the geodesic's coordinate velocity is much less than the speed of light.}
\end{figure}

Fig.~\ref{Massless} shows a plot of $\tilde{r}(\tilde{t})$ for a ball very close to the Buchdahl bound $\tilde{R} = 2/3$.  GR effects are the strongest in this regime, as the ball is on the verge of collapsing into a black hole.  We see that even in this highly relativistic regime, the light ray's coordinate velocity does not change much as it travels through the ball, although it is much smaller than $c$.  But the slight deviation that does occur is rather counterintuitive: the light ray actually appears to \emph{slow down} near the center of the ball, before speeding up again as it escapes.  (In a locally inertial frame, of course, the light ray always appears to be traveling at speed $c$ by the equivalence principle.)  From the perspective of the coordinate frame, the center of the ball appears to repel the light ray!  We give a physical interpretation of this behavior below when we discuss timelike geodesics.  We can calculate its coordinate velocity at the origin by expanding \eqref{masslessroft} to second order in $\tilde{r}$ and $\tilde{t}$, reducing it to
\[
\tilde{r} = \frac{1}{2} \left( 3 \sqrt{1 - 2 \tilde{R}^2} - 1 \right) \tilde{t}.
\]
The light ray's coordinate velocity $v$ at the origin decreases monotonically with $\tilde{R}$ from $v = c$ for $\tilde{R} = 0$ (where the spacetime curvature vanishes) to $v = 0$ for the Buchdahl bound $\tilde{R} = 2/3$.

The case of a massive particle is more complicated, and we will need more sophisticated tools than simply using the metric directly.  For systems with a high degree of symmetry, it is often easier to work with quantities that are conserved under the equations of motion (whose conservation can usually be expressed as a first-order differential equation) than with the equations of motion themselves (which are usually second-order differential equations).  In our case, the metric $\eqref{g}$ is static and has a timelike Killing vector $K := \partial_t$ with components $K^\mu = (1, 0, 0, 0)$, so the quantity
\beq \label{GRE}
\tilde{E} := -\frac{K_\mu U^\mu}{c} = -\frac{g_{tt}}{c} U^t = e^{2 \alpha} \frac{d\tilde{t}}{d\tilde{\tau}},
\eeq
which can be thought of as the total energy divided by the rest energy, is conserved.

[$\tilde{E}$ is also conserved for a massless particle, but in that case $\tau$ must be interpreted as a general affine parameter $\lambda$ rather than as the proper time, and the actual value of $\tilde{E}$ simply sets a choice of affine parameter and has no physical significance.  Indeed, multiplying \eqref{masslessdrdt} and \eqref{GRE} gives $\tilde{E}\, d\tilde{\lambda} / d\tilde{r} = e^{\alpha + \beta}$ (where $\tilde{\lambda} := \omega \lambda$), which can be integrated to
\[
4 \tilde{E} \tilde{\lambda} = 3 \sqrt{2 \big( 1 - 2 \tilde{R}^2 \big)} \arcsin \left( \sqrt{2} \tilde{r} \right) - 2 \tilde{r},
\]
implicitly giving the geodesic's affine parameterization.]

We also have $-U^\mu U_\mu = e^{2 \alpha} \left( U^t \right)^2 - e^{2 \beta} \left( U^r \right)^2 = c^2$ for any timelike trajectory.  We can use \eqref{GRE} to rearrange this equation to
\beq \label{Eeff}
\frac{1}{2} \left( \frac{d\tilde{r}}{d\tilde{\tau}} \right)^2 + \frac{1}{2} e^{-2 \beta} \left( 1 - e^{-2 \alpha} \tilde{E}^2 \right) = 0.
\eeq
Similarly to \eqref{IOM} in Sec.~\ref{Special}, this equation is formally identical to the equation of motion for a nonrelativistic bound particle in one dimension with zero total energy in an effective potential per unit mass
\[
V_\text{eff}(\tilde{r}) = \frac{1}{2} e^{-2 \beta} \left( 1 - e^{-2 \alpha} \tilde{E}^2 \right).
\]
This effective potential is clearly much more complicated than the simple anharmonic potential \eqref{SRVeff} for the SR case.  In particular, unlike \eqref{SRVeff}, the GR effective potential function depends on the ball's total radius $\tilde{R}$ (through $\alpha$) in addition to the particle's total energy $\tilde{E}$, and this dependence has a simple physical effect.  In the case of a ball of uniform charge, Gauss's law (together with spherical symmetry) guarantees that the charged particle feels an electric field whose strenth depends only on the amount of charge at a smaller radius then the particle's current position.  This fact follows directly from Newton's shell theorem: at any given instant, any spherically symmetric static charge outside of a particle's current radius does not affect its motion.  But the shell theorem fails in GR, because the ball's total radius (and therefore the matter that exists all the way out to that radius) affects the metric via $\alpha$ even deep inside the ball: even if we were to release a massive particle from rest at a radius smaller than $\tilde{R}$, the existence of the spherically symmetric static matter outside its complete trajectory would still affect its oscillation period.  One way to physically interpret this fact is that the pressure that the outer mass exerts on the inner mass itself contributes to the stress-energy tensor inside the ball, increasing its effective gravity beyond that due to just its mass density.

\begin{figure}[b]
\includegraphics[width=\columnwidth]{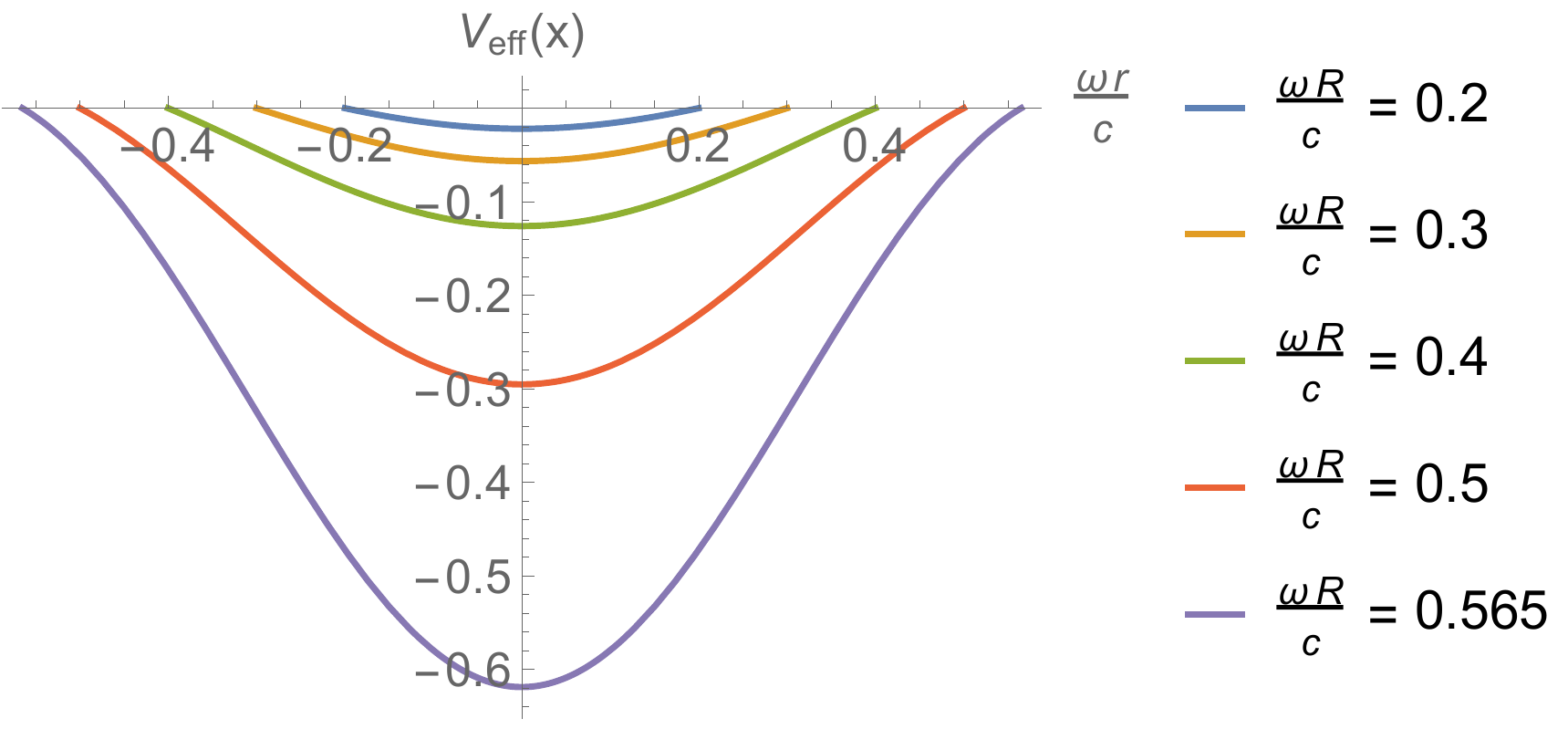}
\caption{\label{GRVeffs} Effective potential \eqref{GRVeff} for various values of $\tilde{R}$.  The inflection points appear at $\tilde{R} = 1/\sqrt{11} = 0.302$.}
\end{figure}

For simplicity, we will only consider the case where the particle's amplitude equals the ball's radius $\tilde{R}$.  $d\tilde{r}/d\tilde{\tau} = 0$ at this turning point, and $V_\text{eff} \big( \tilde{R} \big) = 0$ implies that
\[
\tilde{E} = e^{\alpha \left( \tilde{R} \right)} = \sqrt{1 - 2 \tilde{R}^2} \in (1/3, 1),
\]
where $\tilde{E} \to 1$ in the nonrelativistic limit $\tilde{R} \to 0$, and $\tilde{E} \to 1/3$ in the Buchdahl limit $\tilde{R} \to 2/3$.
The effective potential therefore becomes
\begin{align} \label{GRVeff}
V_\text{eff}(\tilde{r}) =& -\frac{1}{2} \tilde{E}^2\, \frac{x^2 (5 - x) (x-1)}{(3-x)^2}, \\
x :=& \frac{\sqrt{1 - 2 \tilde{r}^2}}{\tilde{E}} = \sqrt{\frac{1 - 2 \tilde{r}^2}{1 - 2 \tilde{R}^2}}. \nonumber
\end{align}
This potential is plotted in Fig.~\ref{GRVeffs}.  It qualitatively resembles the SR effective potential \eqref{SRVeff}, but the inflection points appear at the much lower value of $\tilde{R} = 1/\sqrt{11} = 0.302$.  Just as in the SR case, at the turning point $x = R$ the four-acceleration $d^2x / d\tau^2 = -\omega^2 R$.

\begin{figure}
\includegraphics[width=\columnwidth]{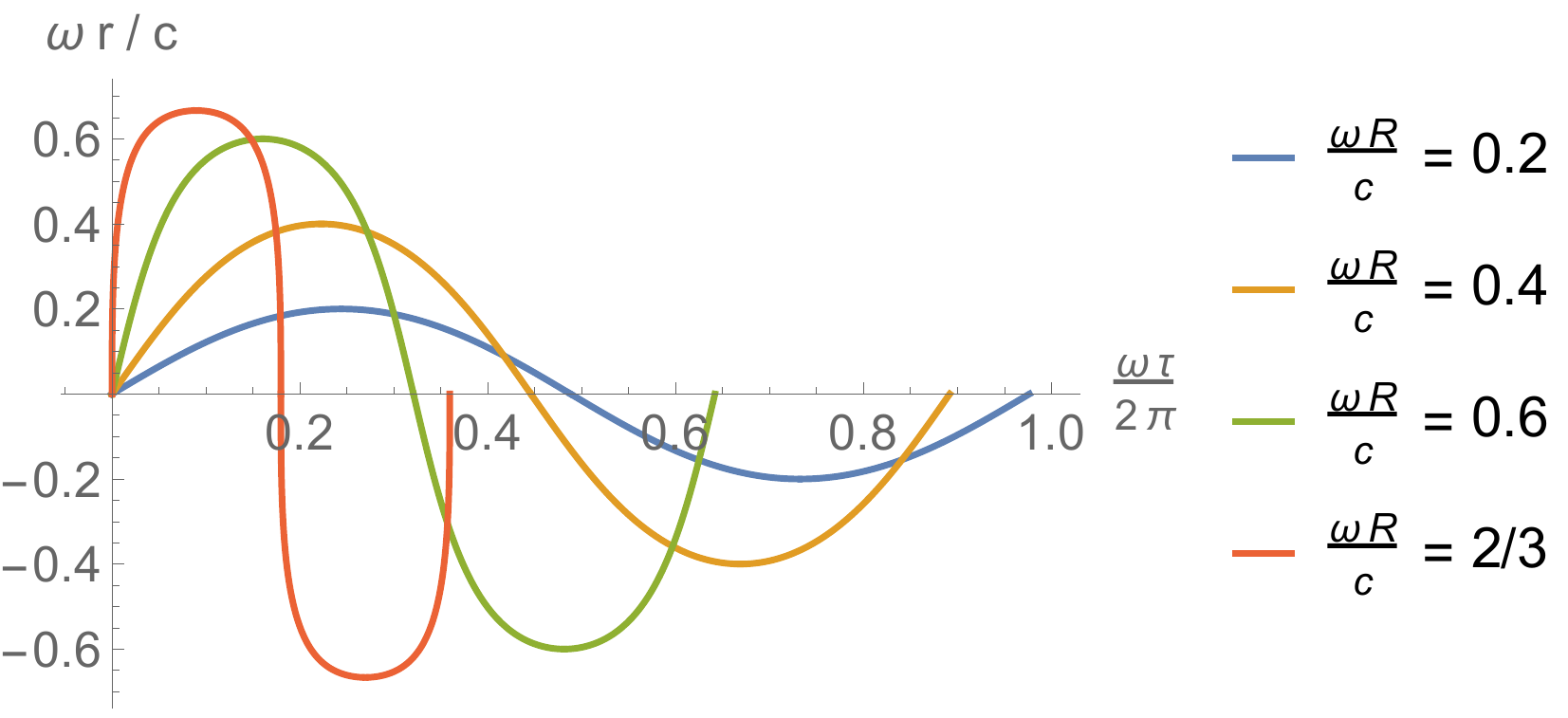}
\caption{\label{rvtau} Timelike geodesic trajectory \eqref{tauofr} plotted against proper time over one complete period.  The red curve corresponds to the maximum value of $\tilde{R}$ allowed by Buchdahl's theorem; its tangent is vertical at the origin and the proper velocity diverges.}
\end{figure}

Solving \eqref{Eeff} for $d\tilde{r}/d\tilde{\tau}$ gives
\begin{align}
\frac{d\tilde{r}}{d\tilde{\tau}} &= \tilde{E} x \frac{\sqrt{(5-x) (x-1)}}{3-x} \label{drdtau} \\
\tilde{\tau} &= \frac{1}{\tilde{E}} \int d\tilde{r} \frac{3-x}{x \sqrt{(5-x) (x-1)}}. \nonumber
\end{align}
While it is not easy (even with software), this integral can be evaluated to
\begin{widetext}
\begin{align}
\tilde{\tau} =\, &\frac{2 \sqrt{2} \tilde{E}}{\sqrt{(\tilde{E} + 1) (5 \tilde{E} - 1)}} \label{tauofr} \\
&\times \left[ F \left( \sqrt{\frac{(5 \tilde{E} - 1) (x - 1)}{(\tilde{E} - 1) (x - 5)}}; \sqrt{\frac{(\tilde{E} - 1) (5 \tilde{E} + 1)}{(\tilde{E} + 1) (5 \tilde{E} - 1)}} \right) - 2 \Pi \left( \sqrt{\frac{\tilde{E} - 1}{5 \tilde{E} - 1}}; \sqrt{\frac{(5 \tilde{E} - 1) (x - 1)}{(\tilde{E} - 1) (x - 5)}}; \sqrt{\frac{(\tilde{E} - 1) (5 \tilde{E} + 1)}{(\tilde{E} + 1) (5 \tilde{E} - 1)}} \right) \right], \nonumber
\end{align}
where $\Pi(x; y; z)$ is the incomplete elliptic integral of the third kind and we have chosen an arbitrary initial condition.  The trajectory $\tilde{r}(\tilde{\tau})$ is plotted in Fig.~\ref{rvtau}; it qualitatively resembles the corresponding SR trajectory plotted in Fig.~\ref{xvtau}, but the flattening is limited by the Buchdahl bound (although the maximum proper velocity remains unbounded).  As in Sec~\ref{Special}, the proper-time period of oscillation
\begin{align}
T_\text{prop} &= \frac{4}{\omega} \left[ \tilde{\tau} \big( \tilde{r} = \tilde{R} \big) - \tilde{\tau} \left( \tilde{r} = 0 \right) \right] \label{GRTprop} \\
&= \frac{8 \sqrt{2} \tilde{E}}{\sqrt{\big( \tilde{E} + 1 \big) \big( 5 \tilde{E} - 1 \big)}} \left[ 2 \Pi \left( \sqrt{\frac{\tilde{E} - 1}{5 \tilde{E} - 1}}; \sqrt{\frac{ \big( \tilde{E} - 1 \big) \big( 5 \tilde{E} + 1 \big)}{\big( \tilde{E} + 1\big) \big( 5 \tilde{E} - 1 \big)}} \right) - K \left( \sqrt{\frac{ \big( \tilde{E} - 1 \big) \big( 5 \tilde{E} +1 \big)}{\big( \tilde{E} + 1 \big) \big( 5 \tilde{E} - 1 \big)}} \right) \right], \nonumber
\end{align}
\end{widetext}
where $\Pi(x; y)$ is the complete elliptic integral of the third kind.  In the nonrelativistic limit $\tilde{R} \ll 1$,
\[
\frac{T_\text{prop}}{T_\text{NR}} = 1 - \frac{9}{16} \tilde{R}^2 + o \left( \tilde{R}^4 \right),
\]
and in the Buchdahl limit $\tilde{R} = 2/3$,
\beq \label{minGRTprop}
\frac{T_\text{prop}}{T_\text{NR}} = \frac{2}{\pi} \left[ 2 \Pi \left( i; \sqrt{2}\, i \right) - K \left( \sqrt{2}\, i \right) \right] = 0.360.
\eeq

\begin{figure}
\includegraphics[width=\columnwidth]{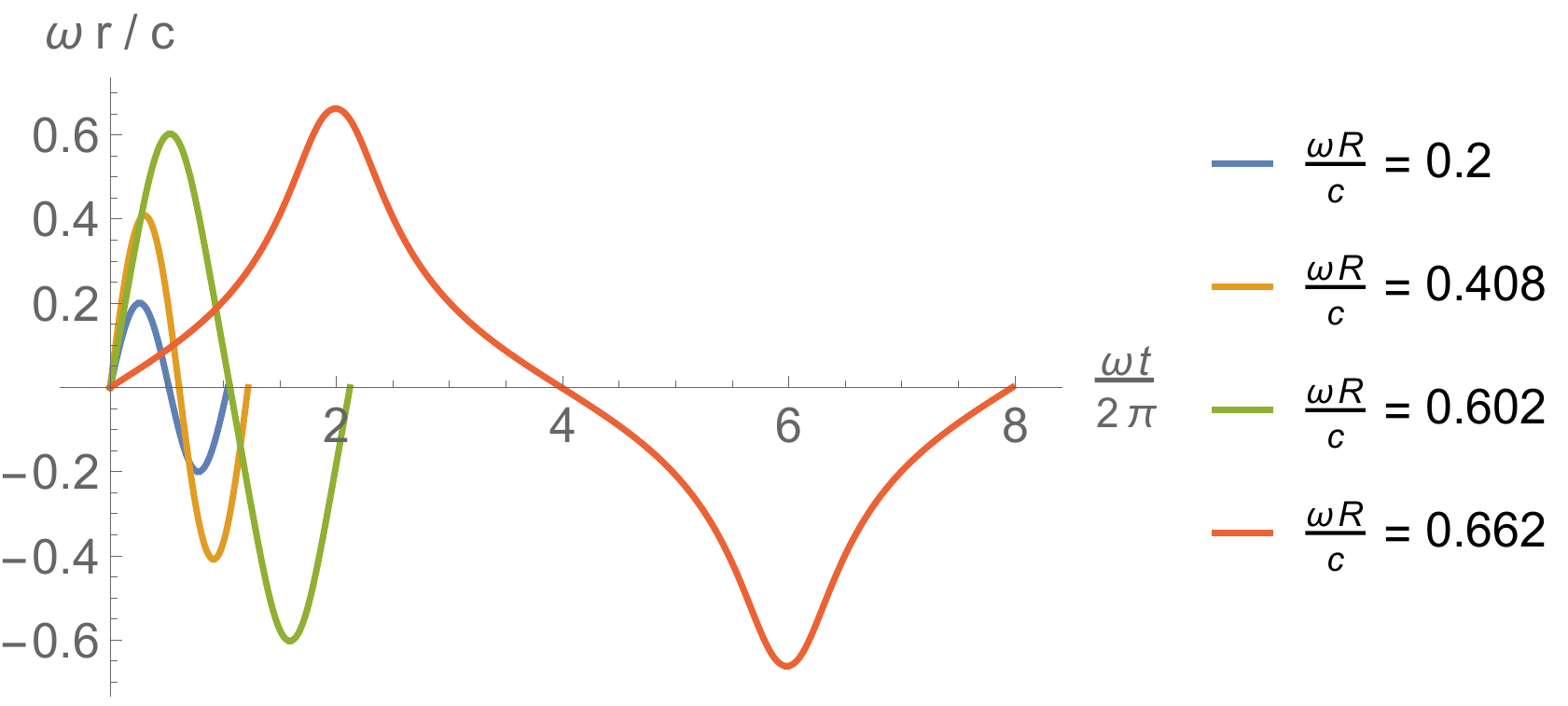}
\caption{\label{rvt} Timelike geodesic trajectory plotted against coordinate time over one complete period.}
\end{figure}

To express the trajectory in terms of coordinate time, we divide \eqref{GRE} by \eqref{drdtau} to get
\beq \label{dtdr}
\frac{d\tilde{t}}{d\tilde{r}} = \frac{4}{\tilde{E}^2 x (3 - x) \sqrt{(5 - x) (x - 1)}}
\eeq
and integrate with respect to $\tilde{r}$ to get $\tilde{t}(\tilde{r})$.  It and the period of oscillation $T_\text{coord}$ are given by the expressions \eqref{tauofr} and \eqref{GRTprop} respectively, but divided by $\tilde{E}$ and with the first argument of the $\Pi$ function multiplied by $i$.  The trajectory $\tilde{r}(\tilde{t})$ is plotted as a function of coordinate time in Fig.~\ref{rvt}.  Just as for the corresponding SR trajectories plotted in Fig.~\ref{xvt}, as $\tilde{R}$ increases, the coordinate period increases, the turning points become sharper, and the coordinate velocity becomes fairly constant throughout the ball's bulk.  But \eqref{dtdr} is a much more complicated function than the corresponding SR expression \eqref{tofx}, and its form qualitatively changes several times as $\tilde{R}$ changes.  For small $\tilde{R}$, the coordinate period
\[
\frac{T_\text{coord}}{T_\text{NR}} = 1 + \frac{15}{16} \tilde{R}^2 + o \left( \tilde{R}^4 \right),
\]
and the coordinate trajectory $\tilde{r}(\tilde{t})$ is approximately sinusoidal.  As $\tilde{R}$ increases from $0$ to $2/3$ and $\tilde{E}$ decreases from $1$ to $1/3$, relativistic effects emerge that qualitatively change the trajectory's behavior.  At $\tilde{R} = 1 / \sqrt{6} = 0.408$ (or equivalently $\tilde{E} = \sqrt{2/3} = .816$), the nondimensionalized inward coordinate acceleration $-d^2\tilde{r} / d\tilde{t}^2$ reaches a maximum value of $\big( \sqrt{2/3} \big) / 3 = 0.272$ and then begins \emph{decreasing} as $\tilde{R}$ increases further.  At $\tilde{R} = 0.469$ (or equivalently $\tilde{E} = 0.748$ is a root of the cubic polynomial $15 \tilde{E}^3 - 9 \tilde{E}^2 - 3 \tilde{E} + 1$), the coordinate speed at the origin similarly begins reaches a maximum value $\tilde{v} = 0.345$ and begins decreasing as $\tilde{R}$ increases further.  At $\tilde{R} = \tilde{E} = 1 / \sqrt{3} = 0.577$, the coordinate acceleration becomes slightly \emph{weaker} at the ball's outer edge than it is just inside the ball's bulk (similarly to the appearance of inflection points in the effective potentials \eqref{SRVeff} and \eqref{GRVeff}).  At $\tilde{R} = 0.602$ (or equivalently $\tilde{E} = 0.525$ is a root of the cubic polynomial $15 \tilde{E}^3 - 37 \tilde{E}^2 + 21 \tilde{E} - 3$), the coordinate velocity $v$ changes from a local maximum to a local minimum at the ball's center, so that in the coordinate frame the particle appears to be slightly repelled from the origin.  Finally, near the Buchdahl limit $\tilde{R} = 2/3$, the particle begins with a small but nonzero (nondimensionalized) inward coordinate acceleration of $2/27= 0.074$, but the coordinate-time period diverges as
\[
\frac{T_\text{coord}}{T_\text{NR}} \sim \sqrt{\frac{2}{\frac{2}{3} - \tilde{R}}},
\]
so in the coordinate frame, a particle released from the ball's outer edge takes arbitrarily long to reach the origin, where its coordinate velocity becomes infinitesimal.

The physical intuition behind the strange coordinate behavior is that near the Buchdahl limit, the pressure at the center of the ball is so high that it is on the verge of collapsing into a black hole.  The fact that the particle takes finite proper time but infinite coordinate time to reach the center of the ball is very reminiscent of the behavior of a timelike geodesic crossing the event horizon of a Schwarzchild black hole.  The apparent (in the coordinate frame) slowing down of the particle as it approaches the center of the ball is merely a result of enormous time dilation -- the particle is not ``truly'' repelled from the center of the ball, any more than infallers are repelled by a black hole just because from far away they appear to slow down as they fall in.  (For example, in our case the geodesic can never reverse direction before reaching the center of the ball.)  Also, near the Buchdahl limit the particle is so relativistic that its mass can be neglected, and its trajectory is very close to null -- and as we saw above, in the coordinate frame null trajectories (not just geodesics) slow down near the ball's center.  Compare the null trajectory plotted in Fig.~\ref{Massless} with the portion near the ball's center of the most relativistic trajectory in Fig.~\ref{rvt}.

\begin{figure}
\includegraphics[width=\columnwidth]{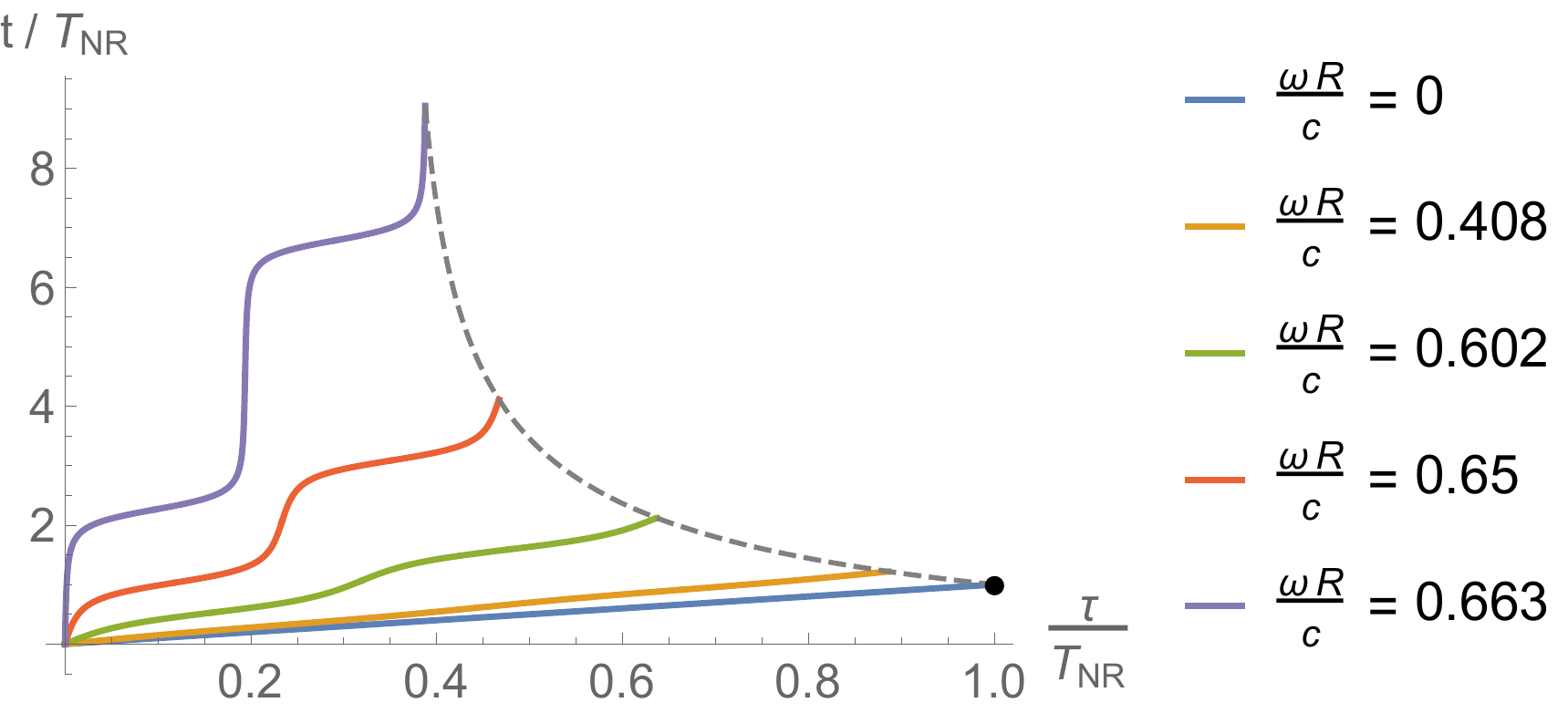}
\caption{\label{GRtvtau} Coordinate time plotted against proper time over one complete oscillation.  The dashed-line envelope parametrically combines the proper-time period \eqref{GRTprop} and the coordinate-time period, and has a vertical asymptote given by \eqref{minGRTprop}.  The black dot indicates the nonrelativistic period.}
\end{figure}

The coordinate time is plotted against the proper time in Fig.~\ref{GRtvtau}.  The curves are qualitatively similar to the SR case shown in Fig.~\ref{tvtau}: in the highly relativistic situation near the Buchdahl bound, a ``staircase'' shape develops as the particle alternates between spending a lot of coordinate time but very little proper time passing through the bulk of the ball (the vertical segments on the plot) and spending a lot of proper time but very little coordinate time near the turning points (the horizontal segments on the plot).  In fact, the discrepancy between the two periods is amplified by the fact that in the coordinate frame, the particle appears to move relatively slowly near the ball's center.  Unlike in the SR case, where the proper-time period can become arbitrarily short, in the GR case it has a minimum value \eqref{minGRTprop}.

For completeness, we will briefly discuss the geodesic equation.  The nonzero Christoffel symbols for the metric \eqref{g} involving the $t$ and $r$ coordinates are \cite{Carroll}
\[
\Gamma^t_{tr} = \Gamma^t_{rt} = \alpha', \qquad \Gamma^r_{tt} = e^{2(\alpha - \beta)} \alpha', \qquad \Gamma^r_{rr} = \beta' \\
\]
where the prime denotes a derivative with respect to $r$.  In our case $\alpha$ and $\beta$ are given by \eqref{alphabeta}, so the geodesic equation becomes
\begin{align}
\frac{d^2t}{d\tau^2} + 2 \alpha' \frac{dr}{d\tau} \frac{dt}{d\tau} &= 0 \label{geo1} \\
\frac{d^2r}{d\tau^2} + c^2 e^{2(\alpha - \beta)} \alpha' \left( \frac{dt}{d\tau} \right)^2 + \beta' \left( \frac{dr}{d\tau} \right)^2 &= 0. \label{geo2}
\end{align}
The first equation is equivalent to the conservation of $\tilde{E}$ as defined in \eqref{GRE}: $d/d\tau (dt / d\tau) = (dr / d\tau)\ d / dr (dt / d\tau)$, so dividing \eqref{geo1} by $dr / d\tau$ gives
\[
\frac{d}{dr} \left( \frac{dt}{d\tau} \right) = -2 \alpha' \frac{dt}{d\tau},
\]
which can be integrated with respect to $r$ to \eqref{GRE} with $\tilde{E}$ as the integration constant.

Note that $d^2r / d\tau^2 = d / dr \left( (1/2) (dr/d\tau)^2 \right)$, which can be seen either from the chain rule or by analogy with the nonrelativistic work-energy theorem $F\, dr = d(\text{KE})$, where $F$ is force and $\text{KE}$ is kinetic energy.  The $r$ component \eqref{geo2} of the geodesic equation can therefore be written as
\[
\left( \frac{d}{dr} + 2 \beta' \right) \left( \frac{1}{2} \left( \frac{dr}{d\tau} \right)^2 \right) = -\tilde{E}^2 c^2\, e^{-2(\alpha + \beta)} \alpha',
\]
which is simply a first-order linear ordinary differential equation for the quantity $(1/2) (dr /d\tau)^2$.  Multiplying by the integrating factor $e^{2 \beta}$ gives
\[
\frac{d}{dr} \left( \frac{1}{2} \left( \frac{dr}{d\tau} \right)^2 e^{2 \beta} \right) = -\tilde{E}^2 c^2\, e^{-2 \alpha} \alpha',
\]
which can be integrated with respect to $r$ to \eqref{Eeff}.  Therefore, the trajectories derived in this section are indeed geodesics.

\section{Conclusion \label{Conclusion}}

Two of physicists' main workhorses for building intuition about nonrelativistic accelerated single-particle systems are the case of uniform acceleration and the simple harmonic oscillator.  The generalization of the former situation to the relativistic context has been extensively studied, but of the latter much less so.  We have found explicit expressions \eqref{xoftau}, \eqref{tofx}, \eqref{masslessroft}, and \eqref{tauofr} relating particles' position, proper time, and coordinate time in two physical setups that naturally generalize the harmonic oscillator into the special- and general-relativisitic regimes.  These relations may help students learning relativity to build intuition about the highly relativistic regime.

The SR and GR situations are qualitatively similar in many ways.  The anharmonic terms in both effective potentials weaken the effective proper force on a test particle far away from the center, to the extent that for sufficiently high particle energies, near the turning points the effective proper force actually \emph{decreases} with increasing radius (although exactly at the turning point $r = R$, it equals the nonrelativistic value $-m \omega^2 R$).  The particle's motion as measured by proper time therefore becomes highly concentrated near the turning points in the ultrarelativistic limit, and its maximum proper velocity becomes arbitrarily large.  On the other hand, the particle cannot exceed the speed of light to cross large diameters, so its motion as measured by coordinate time becomes highly concentrated in the bulk of the ball.  In both cases, the proper-time period of oscillation is less than the nonrelativistic period $2 \pi / \omega$, while the coordinate-time period can become arbitrarily long.

But there are also important differences between the SR and GR cases.  One is that Newton's shell theorem fails in the GR case: spherically symmetric matter outside of the particle's radius affects its trajectory (roughly) by applying gravitating pressure on the inner mass.  Another difference is the emergence of black-hole-like behavior at high pressures: for example, near the center of a sufficiently large ball, the pressure induces such strong time dilation that in the coordinate frame, the particle appears to \emph{slow down} as it approaches the center, just as with the event horizon of a black hole.  A ball of uniform charge can in principle be made arbitarily large, so a test particle can become arbitrarily energetic, but Buchdahl's theorem prohibits a ball of uniform mass from exceeding a radius $\tilde{R} > 2/3$ without collapsing into a black hole.  While this near-black-hole obviously produces some strong relativistic effects, in a sense it also prevents others, by constraining the allowed matter content.  For example, the SR proper-time period $\eqref{SRTprop}$ can become arbitrarily short, but the GR proper-time period \eqref{GRTprop} cannot become shorter than about $36\%$ of the nonrelativistic period $2 \pi / \omega$, as shown in \eqref{minGRTprop}.  Similarly, the SR coordinate velocity at the center of the ball can in principle come arbitrarily close to the speed of light, but in the GR case, the strong time dilation from the near-black-hole prevents the coordinate velocity at the center of the ball from exceeding $0.345 c$.

\bibliography{Biblio}

\begin{thebibliography}{10}%
\makeatletter
\providecommand \@ifxundefined [1]{%
 \@ifx{#1\undefined}
}%
\providecommand \@ifnum [1]{%
 \ifnum #1\expandafter \@firstoftwo
 \else \expandafter \@secondoftwo
 \fi
}%
\providecommand \@ifx [1]{%
 \ifx #1\expandafter \@firstoftwo
 \else \expandafter \@secondoftwo
 \fi
}%
\providecommand \natexlab [1]{#1}%
\providecommand \enquote  [1]{``#1''}%
\providecommand \bibnamefont  [1]{#1}%
\providecommand \bibfnamefont [1]{#1}%
\providecommand \citenamefont [1]{#1}%
\providecommand \href@noop [0]{\@secondoftwo}%
\providecommand \href [0]{\begingroup \@sanitize@url \@href}%
\providecommand \@href[1]{\@@startlink{#1}\@@href}%
\providecommand \@@href[1]{\endgroup#1\@@endlink}%
\providecommand \@sanitize@url [0]{\catcode `\\12\catcode `\$12\catcode
  `\&12\catcode `\#12\catcode `\^12\catcode `\_12\catcode `\%12\relax}%
\providecommand \@@startlink[1]{}%
\providecommand \@@endlink[0]{}%
\providecommand \url  [0]{\begingroup\@sanitize@url \@url }%
\providecommand \@url [1]{\endgroup\@href {#1}{\urlprefix }}%
\providecommand \urlprefix  [0]{URL }%
\providecommand \Eprint [0]{\href }%
\providecommand \doibase [0]{http://dx.doi.org/}%
\providecommand \selectlanguage [0]{\@gobble}%
\providecommand \bibinfo  [0]{\@secondoftwo}%
\providecommand \bibfield  [0]{\@secondoftwo}%
\providecommand \translation [1]{[#1]}%
\providecommand \BibitemOpen [0]{}%
\providecommand \bibitemStop [0]{}%
\providecommand \bibitemNoStop [0]{.\EOS\space}%
\providecommand \EOS [0]{\spacefactor3000\relax}%
\providecommand \BibitemShut  [1]{\csname bibitem#1\endcsname}%
\let\auto@bib@innerbib\@empty
\bibitem [{\citenamefont {Kleppner}\ and\ \citenamefont {Kolenkow}(1973)}]{KK}%
  \BibitemOpen
  \bibfield  {author} {\bibinfo {author} {\bibfnamefont {D.}~\bibnamefont
  {Kleppner}}\ and\ \bibinfo {author} {\bibfnamefont {R.~J.}\ \bibnamefont
  {Kolenkow}},\ }\href@noop {} {\emph {\bibinfo {title} {An Introduction to
  Mechanics}}}\ (\bibinfo  {publisher} {McGraw-Hill},\ \bibinfo {year} {1973})\
  \bibinfo {note} {problem 2.26}\BibitemShut {NoStop}%
\bibitem [{\citenamefont {Cooper}(1966)}]{Cooper}%
  \BibitemOpen
  \bibfield  {author} {\bibinfo {author} {\bibfnamefont {P.~W.}\ \bibnamefont
  {Cooper}},\ }\href {\doibase 10.1119/1.1972773} {\bibfield  {journal}
  {\bibinfo  {journal} {American Journal of Physics}\ }\textbf {\bibinfo
  {volume} {34}},\ \bibinfo {pages} {68} (\bibinfo {year} {1966})}\BibitemShut
  {NoStop}%
\bibitem [{\citenamefont {Jackson}(1999)}]{Jackson}%
  \BibitemOpen
  \bibfield  {author} {\bibinfo {author} {\bibfnamefont {J.~D.}\ \bibnamefont
  {Jackson}},\ }\href@noop {} {\emph {\bibinfo {title} {Classical
  Electrodynamics}}},\ \bibinfo {edition} {3rd}\ ed.\ (\bibinfo  {publisher}
  {Wiley},\ \bibinfo {year} {1999})\ \bibinfo {note} {section 11.9}\BibitemShut
  {NoStop}%
\bibitem [{\citenamefont {Moreau}\ \emph {et~al.}(1994)\citenamefont {Moreau},
  \citenamefont {Easther},\ and\ \citenamefont {Neutze}}]{Moreau}%
  \BibitemOpen
  \bibfield  {author} {\bibinfo {author} {\bibfnamefont {W.}~\bibnamefont
  {Moreau}}, \bibinfo {author} {\bibfnamefont {R.}~\bibnamefont {Easther}}, \
  and\ \bibinfo {author} {\bibfnamefont {R.}~\bibnamefont {Neutze}},\
  }\href@noop {} {\bibfield  {journal} {\bibinfo  {journal} {American Journal
  of Physics}\ }\textbf {\bibinfo {volume} {62}},\ \bibinfo {pages} {531}
  (\bibinfo {year} {1994})}\BibitemShut {NoStop}%
\bibitem [{\citenamefont {Harvey}(1972)}]{Harvey}%
  \BibitemOpen
  \bibfield  {author} {\bibinfo {author} {\bibfnamefont {A.~L.}\ \bibnamefont
  {Harvey}},\ }\href {\doibase 10.1103/PhysRevD.6.1474} {\bibfield  {journal}
  {\bibinfo  {journal} {Phys. Rev. D}\ }\textbf {\bibinfo {volume} {6}},\
  \bibinfo {pages} {1474} (\bibinfo {year} {1972})}\BibitemShut {NoStop}%
\bibitem [{\citenamefont {Babusci}\ \emph {et~al.}(2013)\citenamefont
  {Babusci}, \citenamefont {Dattoli}, \citenamefont {Quattromini},\ and\
  \citenamefont {Sabia}}]{Babusci}%
  \BibitemOpen
  \bibfield  {author} {\bibinfo {author} {\bibfnamefont {D.}~\bibnamefont
  {Babusci}}, \bibinfo {author} {\bibfnamefont {G.}~\bibnamefont {Dattoli}},
  \bibinfo {author} {\bibfnamefont {M.}~\bibnamefont {Quattromini}}, \ and\
  \bibinfo {author} {\bibfnamefont {E.}~\bibnamefont {Sabia}},\ }\href
  {\doibase 10.1103/PhysRevE.87.033202} {\bibfield  {journal} {\bibinfo
  {journal} {Phys. Rev. E}\ }\textbf {\bibinfo {volume} {87}},\ \bibinfo
  {pages} {033202} (\bibinfo {year} {2013})}\BibitemShut {NoStop}%
\bibitem [{\citenamefont {de~la Fuente}\ and\ \citenamefont
  {Torres}(2016)}]{delaFuente}%
  \BibitemOpen
  \bibfield  {author} {\bibinfo {author} {\bibfnamefont {D.}~\bibnamefont
  {de~la Fuente}}\ and\ \bibinfo {author} {\bibfnamefont {P.~J.}\ \bibnamefont
  {Torres}},\ }\href@noop {} {\bibfield  {journal} {\bibinfo  {journal}
  {Qualitative Theory of Dynamical Systems}\ ,\ \bibinfo {pages} {1}} (\bibinfo
  {year} {2016})}\BibitemShut {NoStop}%
\bibitem [{\citenamefont {Goldstein}\ \emph {et~al.}(2002)\citenamefont
  {Goldstein}, \citenamefont {Poole},\ and\ \citenamefont {Safko}}]{Goldstein}%
  \BibitemOpen
  \bibfield  {author} {\bibinfo {author} {\bibfnamefont {H.}~\bibnamefont
  {Goldstein}}, \bibinfo {author} {\bibfnamefont {C.}~\bibnamefont {Poole}}, \
  and\ \bibinfo {author} {\bibfnamefont {J.}~\bibnamefont {Safko}},\
  }\href@noop {} {\emph {\bibinfo {title} {Classical Mechanics}}},\ \bibinfo
  {edition} {3rd}\ ed.\ (\bibinfo  {publisher} {Addison Wesley},\ \bibinfo
  {year} {2002})\ \bibinfo {note} {sections 7.9-10}\BibitemShut {NoStop}%
\bibitem [{\citenamefont {Urabe}(1962)}]{Urabe}%
  \BibitemOpen
  \bibfield  {author} {\bibinfo {author} {\bibfnamefont {M.}~\bibnamefont
  {Urabe}},\ }\href {http://projecteuclid.org/euclid.hmj/1206139730} {\bibfield
   {journal} {\bibinfo  {journal} {J. Sci. Hiroshima Univ. Ser. A-I Math.}\
  }\textbf {\bibinfo {volume} {26}},\ \bibinfo {pages} {93} (\bibinfo {year}
  {1962})}\BibitemShut {NoStop}%
\bibitem [{\citenamefont {Carroll}(2004)}]{Carroll}%
  \BibitemOpen
  \bibfield  {author} {\bibinfo {author} {\bibfnamefont {S.~M.}\ \bibnamefont
  {Carroll}},\ }\href@noop {} {\emph {\bibinfo {title} {Spacetime and Geometry:
  An Introduction to General Relativity}}}\ (\bibinfo  {publisher} {Pearson},\
  \bibinfo {year} {2004})\ \bibinfo {note} {chapter 5}\BibitemShut {NoStop}%
\end{thebibliography}%
\bibliographystyle{apsrev4-1}

\end{document}